\documentclass[aps,prb,twocolumn,superscriptaddress,longbibliography,floatfix]{revtex4-1}

\usepackage[T1]{fontenc}
\usepackage[utf8]{inputenc}
\usepackage{lmodern}
\usepackage{microtype}
\usepackage{amsmath,amssymb,bm}
\usepackage{graphicx}
\usepackage{booktabs}
\usepackage{siunitx}
\usepackage{physics}
\usepackage{xcolor}
\usepackage{enumitem}
\usepackage{hyperref}
\usepackage{enumitem}

\newcommand{\doilink}[1]{\href{https://doi.org/#1}{doi:#1}}

\begin{document}

\title{An iterative method bridging DFT, disorder averaging, and experiment in intercalated materials: application to Au-intercalated graphene}
\author{Poonam Kumari}
\affiliation{Universit\'e Paris Saclay, CNRS, CEA, Institut de Physique Th\'eorique, 91191, Gif-sur-Yvette, France}
\author{Alberto Zobelli}
\affiliation{Universit\'e Paris Saclay, CNRS, Laboratoire de Physique des Solides, Orsay, France}
\author{Igor de Melo Froldi}
\affiliation{Universit\'e Paris Saclay, CNRS, CEA, Institut de Physique Th\'eorique, 91191, Gif-sur-Yvette, France}
\affiliation{Instituto de F{\'i}sica, Universidade Federal de Goi{\'a}s, 74.690-900, Goi{\^a}nia-GO, Brazil}
\author{Adeline Cr\'epieux}
	\affiliation{Aix Marseille Univ, Universit\'e de Toulon, CNRS, CPT, Marseille, France}
\author{Laurent Simon}
\affiliation{Institut de Sciences des Mat\'eriaux de Mulhouse, CNRS-UMR 7361
Universit\'e de Haute Alsace, Mulhouse 68093, France}
\author{Cristina Bena}
\affiliation{Universit\'e Paris Saclay, CNRS, CEA, Institut de Physique Th\'eorique, 91191, Gif-sur-Yvette, France}
\date{\today}

\begin{abstract}
Intercalation can strongly modify the electronic dispersion of a host material, as directly revealed by angle-resolved photoemission spectroscopy (ARPES). We develop a general iterative method combining density functional theory (DFT), tight-binding (TB), disorder averaging within the self-consistent T-matrix approximation (SCTMA), and experiment, to construct an effective model of the intercalated system. DFT identifies the relevant microscopic degrees of freedom and constrains selected model parameters, while comparison of SCTMA calculations with experiment guides their further refinement. We apply this method to graphene intercalated with Au clusters~\cite{NairPRB2012} and show that it reproduces the main ARPES signatures of the Au-cluster phase, including the broadening of the van Hove singularity and the emergence of kink-like features in the dispersion. The essential microscopic ingredients identified by the analysis are the hybridization between selected intercalant orbitals and the graphene states, together with an intercalation-induced local scattering potential.
\end{abstract}

\maketitle

\section{Introduction}
\label{sec:intro}

The electronic structure of two-dimensional van der Waals materials can be
engineered by both geometric and chemical means. Moir\'e superlattices, obtained
for example by twisting adjacent graphene layers, provide a geometric route to
narrow or nearly flat bands and to correlated insulating and superconducting
states~\cite{Bistritzer2011Moire,Cao2018Correlated,Cao2018SC}.
Intercalation---the insertion of atoms or compounds between graphene and a
substrate, or between adjacent graphene layers---provides a complementary
chemical route. In epitaxial graphene on SiC or metallic substrates,
intercalation can break or screen graphene--substrate bonds and restore
quasi-free-standing $\pi$ bands, while charge transfer and interface dipoles
control the carrier density~\cite{Riedl2009H,Starke2012Intercalation,Briggs2019IntercalationReview,
Dedkov2017Intercalation,Gierz2010Au,Emtsev2011Ge}.
Ordered intercalant layers can additionally impose superlattice potentials, modify sublattice and spin--orbit couplings, and stabilize atomically thin intercalant phases protected by the graphene overlayer~\cite{Caffrey2016Li,Forti2020Au2D,Warmuth2016Bi}.
Interestingly enough, alkali-metal and rare-earth intercalation can generate strong electron doping,
allowing the Fermi level to approach or even cross the graphene van Hove
singularity and the associated Lifshitz transition~\cite{Rosenzweig2019Yb,Rosenzweig2020Overdoping,Zaarour2023Er}.

Interpreting and modeling the effective band structure of intercalated graphene from ARPES measurements is not always straightforward. ARPES probes the spatially-averaged momentum-resolved spectral function~\cite{Sobota2021ARPES}, and is therefore intrinsically sensitive to spatial variations and disorder induced by intercalation.
The spectral function may exhibit features such as kinks, 
due to an avoided crossing or hybridization between the graphene
C-$p_z$ bands and the substrate or intercalant orbitals~\cite{Gruneis2008Hybridization}. Heavy-element intercalants can similarly
produce hybridization gaps, symmetry-breaking gaps, or sizeable spin--orbit
splittings~\cite{Marchenko2012AuSOC,Warmuth2016Bi}. At still higher doping, the graphene
dispersion around the $M$ point may be strongly reshaped, producing extended
van Hove features, Lifshitz transitions, and flat-band-like states~\cite{NairPRB2012,Rosenzweig2019Yb,Rosenzweig2020Overdoping,Zaarour2023Er}.

For graphene with spatially disordered intercalants, it remains challenging to combine a first-principles description of the local electronic structure with large-scale configurational disorder in a single effective tight-binding (TB) framework capable of reproducing the experimental observations.

To solve this problem, we construct a general iterative procedure that combines density functional theory (DFT), TB, disorder averaging within the self-consistent T-matrix approximation (SCTMA), and comparison with experiments. This procedure allows us to write down an effective model for the intercalated system. The DFT calculations identify the relevant microscopic degrees of freedom and constrain part of the TB parameter space, while a comparison between the SCTMA spectra and the experimental data is used to refine the remaining parameters. The procedure allows one to calculate the disorder-averaged momentum-resolved spectral function and the effective band-structure of the intercalated system.

\begin{figure}[t]
  \centering
  \includegraphics[width=0.7\columnwidth, angle=0]{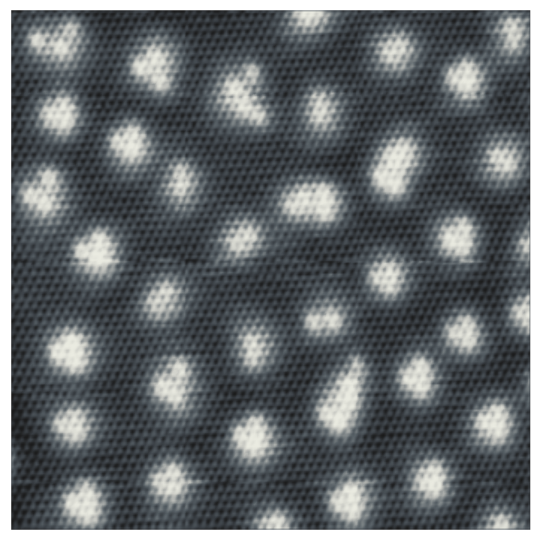}
  \caption{Experimental image of the cluster phase}
  \label{fig:cluster_phase}
\end{figure}

We apply this technique to Au intercalation beneath epitaxial graphene on SiC. This is known to produce multiple structural phases~\cite{PremlalAPL2009}, including a ``cluster'' phase often referred to as ``ostrich leather' (see Fig.~\ref{fig:cluster_phase}). In this phase ARPES reveals a striking reconstruction of the occupied graphene dispersion near the Van Hove singularity (VHS): the saddle-point region broadens and develops kinks, while the Dirac-point doping is shifted only slightly~\cite{NairPRB2012}. We show that these observed experimental features can be reproduced by our procedure.

Sec.~\ref{sec:methodology_summary} summarizes the computational workflow. Sec.~\ref{sec:methodology} presents the theoretical models and approaches: DFT and Wannierization, the exact periodic TB formalism, and the SCTMA formalism. Sec.~\ref{sec:pristine} establishes the pristine-graphene reference, Secs.~\ref{sec:dft-i}--\ref{sec:effective_tb} present the Au-intercalated DFT/Wannier results and effective-model parameters, and Sec.~\ref{sec:sctma} compares the disorder-averaged spectral function with experiment. We conclude in section VI.

\section{Methodology overview}
\label{sec:methodology_summary}
\begin{enumerate}[label=\roman*) ]
    \item {\bf DFT.} We start by performing a periodic DFT supercell calculation of hollow-site Au intercalants. We consider a supercell size consistent with the experimental Au density. We unfold and graphene-project the supercell dispersion to the primitive Brillouin zone to obtain the effective graphene band structure. 
    
\item {\bf Wannierization, filtering and truncation.} We perform a Wannierization to obtain the TB hopping parameters between the Au and C atomic orbitals. We restrict the number of orbitals and the range of hopping to a maximum to reproduce the DFT results in the relevant energy window. Thus we find that pristine graphene is accurately described by a $p_z$ six-nearest-neighbor (6NN) TB model. We show that only a small subset of Au orbitals is relevant, in particular the $d_{xz}/d_{yz}$ and $d_{x^2-y^2}/d_{xy}$, for which the hybridization is short ranged and is well captured by the coupling of each Au orbital to the six nearest carbon atoms around a hollow site. The procedure also provides the symmetry label $m_{\alpha}$ associated with each Au orbital $\alpha$. We check via a TB exact calculation that this truncation and filtering reproduces the exact DFT spectra.

\item {\bf TB effective model.} This procedure provides us with the first iteration of the TB parameters in the model. However it neglects various important factors such as the SiC substrate, the buffer layer, and the random distribution of the intercalants. Thus it cannot correctly model the electrostatics of the graphene-intercalant interaction and of the intercalant-substrate interactions, nor the exact distance between the intercalant and graphene. We thus use it a controlled extraction tool for:
\begin{enumerate}[label=\alph*)]
\item The orbitals that hybridize with graphene in the experimental energy window  
\item Their symmetry channel coupling 
\item The hopping range and order of magnitude for the hopping amplitudes between the intercalant and graphene. 
\end{enumerate}
It will not provide 
\begin{enumerate}[resume]
\item The exact energies of the intercalant orbital levels
\item The electrostatic potential seen by the C atoms in the vicinity of the intercalant 
\item The exact C-intercalant hybridization value. 
\end{enumerate}
We will thus take some phenomenological values for the d)-f) parameters fine-tuned to provide the best agreement with the experimental data. With the exception of the on-site electrostatic potential, all parameter values lie within the physically relevant ranges obtained from DFT.

\item {\bf SCTMA.}
We implement this information into an SCTMA calculation for randomly distributed intercalants, and we compute the effective spectral function of intercalated graphene. We take the effective intercalant density to be similar to that of the experimental samples

 \item {\bf Experimental comparison and fine-tuning of parameters.} We compare the resulting spectral function with the one measured experimentally by ARPES~\cite{NairPRB2012}. We adjust the d)-f) parameters above to best fit the experimental data. 
Within this framework the SCTMA calculations reproduce very well the kink-like renormalizations and an extension of the VHS region, in agreement with ARPES measurements of the cluster phase~\cite{NairPRB2012}.

\item {\bf Effective TB model.} Using the parameters determined above, we construct the final effective tight-binding model of the intercalated system.

\end{enumerate}

We therefore propose that dilute intercalated systems can be described quantitatively by an effective TB model constructed through an iterative analysis combining DFT, SCTMA, and experiment. The method is broadly applicable to systems with sparse, disordered intercalants and may be particularly useful for materials in which intercalation gives rise to flat bands~\cite{NairPRB2012,Rosenzweig2019Yb,Rosenzweig2020Overdoping,Zaarour2023Er}, with potential relevance to the search for high-temperature superconductivity.

\section{Theoretical framework}
\label{sec:methodology}

\subsection{DFT and Wannierization}
\label{sec:dft}

The DFT calculations were performed using both OpenMX and Quantum ESPRESSO (QE)~\cite{Giannozzi2009QE,Giannozzi2017QE}. The pristine graphene band structure and the Au-intercalated graphene supercell were first calculated with OpenMX using a linear combination of pseudoatomic orbitals (LCPAO) basis~\cite{OpenMXManualUnfolding,Rubel2023Unfolding}. Exchange and correlation effects were treated within the generalized gradient approximation using the Perdew--Burke--Ernzerhof (PBE) functional. Norm-conserving pseudopotentials from the OpenMX DFT$\_$DATA19 database were employed to describe the electron--ion interactions, together with the C6.0-s2p2d1 and Au7.0-s3p2d2f1 pseudoatomic-orbital basis sets for C and Au, respectively. A vacuum layer of $17.6~\AA$ was introduced to suppress interactions between periodically repeated slabs, and an energy cutoff of 150 Ry was used. The Brillouin zone was sampled using $64\times64\times1$ and $16\times16\times1$ $\mathbf{k}$-point meshes for pristine graphene and the Au-intercalated $5\times5$ graphene supercell, respectively. Band unfolding was performed over a uniform reciprocal-space mesh following the procedure described in Refs.~\onlinecite{OpenMXManualUnfolding,Ku2010Unfolding}.

To construct the Wannier representation, equivalent calculations were carried out with QE interfaced with Wannier90~\cite{Wannier1937,Mostofi2008Wannier90}. This implementation allowed us to benchmark the resulting graphene Hamiltonian against previously reported TB parametrizations~\cite{ReichPRB2002,TranAIPAdv2017,KunduMPLB2011,JungMacDonald2013,FangKaxiras2016}. The QE calculations employed a plane-wave basis, scalar-relativistic ultrasoft pseudopotentials, and the PBE exchange-correlation functional. The Brillouin zone was sampled using $64\times64\times1$ and $8\times8\times1$ $\mathbf{k}$-point meshes for pristine graphene and the Au-intercalated $5\times5$ supercell, respectively. Effective Hamiltonians were subsequently obtained from maximally localized Wannier functions using C $p$ orbitals and Au $s$, $p$, and $d$ orbitals as the initial projections.

The resulting electronic structure was verified to be consistent with the OpenMX calculations within the energy range relevant to the present study.

\subsection{TB model}

\subsubsection{Pristine graphene}
\label{pristine}
Graphene is described by a six-neighbor TB Hamiltonian on the honeycomb lattice. 
We use a two-component basis
\[
  \Psi_{\bf k} =
  \begin{pmatrix}
    c_{A{\bf k}} \\ c_{B{\bf k}}
  \end{pmatrix},
\]
where $A$ and $B$ denote the two carbon sublattices. 

In this basis, the graphene Hamiltonian can be written as:
\begin{equation}
  H_0({\bf k})
  =
  \begin{pmatrix}
    \epsilon_{\rm same}({\bf k})-\mu & h_{\rm AB}({\bf k}) \\
    h_{\rm AB}^*({\bf k}) & \epsilon_{\rm same}({\bf k})-\mu
  \end{pmatrix},
  \label{eq:H0graphene6NN}
\end{equation}
with
\begin{align}
  h_{\rm AB}({\bf k})
  &=
  t_1 f_1({\bf k})
  +
  t_3 f_3({\bf k})
  +
  t_4 f_4({\bf k}),
  \label{eq:hAB6NNsimple}
  \\
  \epsilon_{\rm same}({\bf k})
  &=
  t_2 g_2({\bf k})
  +
  t_5 g_5({\bf k})
  +
  t_6 g_6({\bf k}) .
  \label{eq:epsSame6NNsimple}
\end{align}
Here the functions \(f_n({\bf k})\) describe the hoppings between opposite
sublattices, while \(g_n({\bf k})\) describe the hoppings between sites on the same
sublattice:
\begin{align}
  f_n({\bf k})
  &=
  \sum_{\bm{\delta}\in S_n^{AB}}
  e^{i{\bf k}\cdot\bm{\delta}},
  \qquad n=1,3,4,
  \label{eq:fnAB}
  \\
  g_n({\bf k})
  &=
  \sum_{\bm{\rho}\in S_n^{AA}}
  e^{i{\bf k}\cdot\bm{\rho}},
  \qquad n=2,5,6 .
  \label{eq:gnAA}
\end{align}
where
\begin{align}
  S_1^{AB}
  &=\{\boldsymbol{\delta}_1,\boldsymbol{\delta}_2,\boldsymbol{\delta}_3\}\nonumber\\
    S_2^{AA}
  &=
  \left\{
  \pm {\bf a}_1,
  \pm {\bf a}_2,
  \pm({\bf a}_2-{\bf a}_1)
  \right\},
  \nonumber\\
  S_3^{AB}
  &=\{-2\boldsymbol{\delta}_1,-2\boldsymbol{\delta}_2,-2\boldsymbol{\delta}_3\},
\nonumber\\
  S_4^{AB}
  &=
  \left\{
  2\boldsymbol{\delta}_i-\boldsymbol{\delta}_j
  \;,\;
  i,j\in\{1,2,3\},\ i\neq j
  \right\},
  \nonumber\\
  S_5^{AA}
  &=
  \left\{
  \pm({\bf a}_1+{\bf a}_2),
  \pm(2{\bf a}_1-{\bf a}_2),
  \pm({\bf a}_1-2{\bf a}_2)
  \right\},
  \nonumber\\
  S_6^{AA}
  &=
  \left\{
  \pm2{\bf a}_1,
  \pm2{\bf a}_2,
  \pm2({\bf a}_2-{\bf a}_1)
  \right\}.
  \label{eq:longerRangeNeighborShells}
\end{align}
where the graphene Bravais-lattice vectors,
\begin{equation}
  {\bf a}_1=\boldsymbol{\delta}_1-\boldsymbol{\delta}_2,
  \qquad
  {\bf a}_2=\boldsymbol{\delta}_1-\boldsymbol{\delta}_3.
\end{equation}
and the nearest-neighbor vectors
are chosen as
\begin{equation}
  \boldsymbol{\delta}_1 = a(0,1),   \boldsymbol{\delta}_2 = a\left(\frac{\sqrt{3}}{2},-\frac{1}{2}\right),   \boldsymbol{\delta}_3 = a\left(-\frac{\sqrt{3}}{2},-\frac{1}{2}\right),
\end{equation}
where $a$ is the carbon--carbon distance. 

The shells $S_2^{AA}$, $S_4^{AB}$, $S_5^{AA}$, and $S_6^{AA}$ each contain
six vectors, of respective lengths $\sqrt{3}a$, $\sqrt{7}a$, $3a$, and
$2\sqrt{3}a$. The
superscripts $AA$ and $AB$ indicate whether the corresponding hopping connects
equal or opposite graphene sublattices. Since the same-sublattice shells occur
in opposite pairs, $g_2({\bf k})$, $g_5({\bf k})$, and $g_6({\bf k})$ are real.

We use the values of the hopping parameters as obtained from the Wannierization of the DFT calculations:
\begin{eqnarray}
  t_1 &=& -2.937\,{\rm eV}\nonumber \\
  t_2 &=& 0.249\,{\rm eV}\nonumber\\   t_3 &=& -0.260\,{\rm eV}\nonumber\\
  t_4 &=& 0.025\,{\rm eV}\nonumber\\  t_5 &=& 0.050\,{\rm eV}\nonumber\\ t_6&=&-0.024\,{\rm eV}.
  \label{eq:TBparams}
\end{eqnarray}

The longer-range hoppings break particle-hole symmetry and shift the Dirac
point of the bare tight-binding Hamiltonian away from zero energy. We
therefore choose
\begin{equation}
\mu=-0.377~{\rm eV},
\end{equation}
so that the Dirac point coincides with the Fermi level and the pristine
system remains at charge neutrality. Thus, the finite numerical value of
$\mu$ reflects the chosen energy reference and does not by itself describe
doping. Physical doping is instead determined by the displacement of the
Fermi level relative to the Dirac-point energy.

The values of some of the parameters remain consistent with previous references~\cite{ReichPRB2002,TranAIPAdv2017,KunduMPLB2011,JungMacDonald2013,FangKaxiras2016}. Incidentally, note that taking into account hopping up to the sixth nearest-neighbor flips the sign of $t_2$ compared to a three nearest-neighbour parametrization.

The retarded Green's function of pristine graphene is
\begin{equation}
  G_0^R({\bf k},\omega)
  =
  \left[
    (\omega+i\eta)\mathbb{I}_2   -
    H_0({\bf k})
  \right]^{-1},
  \label{eq:G0graphene}
\end{equation}
where $\eta$ is a phenomenological broadening. In the numerical plots below we use
\begin{equation}
  \eta = 0.17\,{\rm eV}.
\end{equation}
This value is taken to make correspondence to the linewidth observed in the ARPES measurements~\cite{NairPRB2012}. The total spectral function is given by
\begin{equation}
  A_0({\bf k},\omega)
  =
  -\frac{1}{\pi}\,
  {\rm Im}\,
  {\rm Tr}\,
  G_0^R({\bf k},\omega).
  \label{eq:A0graphene}
\end{equation}

\subsubsection{Au--C coupling}

To connect the carbon-resolved Wannier couplings to the symmetry-adapted
impurity model, we associate each Au orbital $\alpha$ with a ring-symmetry
label $m_{\alpha}$ and define the corresponding normalized ring harmonic
\begin{equation}
  c_{m_{\alpha}}
  =
  \frac{1}{\sqrt{6}}
  \sum_{j=1}^{6}
  e^{-i m_{\alpha}\theta_j}c_j.
  \label{eq:ringharmonics}
\end{equation}
where $c_j$ annihilates an electron on the $j$ site of the graphene ring around the Au impurity.
For an Au orbital $\alpha$, the Wannierization provides the six-component
coupling vector
\begin{equation}
  \bm{V}_{\alpha}
  =
  \left(
    V_{1\alpha},
    \ldots,
    V_{6\alpha}
  \right)^{T}.
\end{equation}
The corresponding microscopic hybridization is
\begin{equation}
  \hat{H}_{\rm hyb}^{(\alpha)}
  =
  d_{\alpha}^{\dagger}
  \sum_{j=1}^{6}
  V_{j\alpha}c_j
  +
  {\rm h.c.}
  \label{eq:hybrealspace}
\end{equation}
where $d_{\alpha}$ annihilates an electron in Au orbital $\alpha$.
This can equivalently be expressed in the ring-harmonic basis as
\begin{equation}
  \hat{H}_{\rm hyb}^{(\alpha)}
  =
  t_{\alpha} d_{\alpha}^{\dagger}c_{m_{\alpha}}
  +
  {\rm h.c.},
  \label{eq:hybringchannels_hamiltonian}
\end{equation}
where
\begin{equation}
  t_{\alpha}
  =
  \frac{1}{\sqrt{6}}
  \sum_{j=1}^{6}
  V_{j\alpha}e^{i m_{\alpha}\theta_j}.
  \label{eq:hybringchannels}
\end{equation}
This projection provides the direct connection between the Wannier
Hamiltonian and the compact impurity model used in the SCTMA calculation. The quantity
$|t_{\alpha}|$ is the total symmetry-projected hybridization between
the Au orbital $\alpha$ and the corresponding six-site graphene ring
harmonic $c_{m_{\alpha}}$. 

The Au impurities also introduce an electrostatic local on-site potential $U$ on the six atoms in the nearby ring. \begin{equation}
  \hat{H}_U=U\sum_{j=1}^6
  c_{j}^\dagger c_{j}
  \end{equation}
  This plays also the role of local scattering potential.

\subsection{Periodic TB formalism}
\label{sec:periodic_tb}

For a perfectly periodic array of intercalants, the electronic structure can be obtained exactly using a TB formalism. The experimental intercalant distribution is, however, spatially disordered, so a periodic model cannot capture configurational averaging, disorder-induced broadening, or the resulting effective band structure. We therefore use the periodic calculation only as a reference and as a consistency check of the truncated Wannier Hamiltonian.

For this periodic model, we consider a $5\times5$ graphene supercell with lattice vectors:
$\bm{A}_i=5\bm{a}_i$ and reciprocal vectors
$\bm{B}_i\cdot\bm{A}_j=2\pi\delta_{ij}$. The supercell contains
$N_C=50$ carbon $p_z$ orbitals and one hollow-site Au impurity. We denote the
retained Au orbitals by $\alpha\in\mathcal{A}$ and their associated
ring-symmetry labels by $m_{\alpha}\in\{+1,-1,+2,-2\}$. The operators
$c_{i\bm{k}}$ and $d_{\alpha\bm{k}}$ annihilate, respectively, a carbon
electron at site $i$ and an Au electron in orbital $\alpha$ at supercell
momentum $\bm{k}$. The periodic Hamiltonian is
\begin{equation}
  \hat{H}_{\rm per}(\bm{k})
  =
  \hat{H}_C(\bm{k})
  +
  \hat{H}_d(\bm{k})
  +
  \hat{H}_{\rm hyb}(\bm{k})
  +
  \hat{H}_U(\bm{k}).
  \label{eq:periodic_second_quantized_H}
\end{equation}

The graphene contribution is
\begin{equation}
  \hat{H}_C(\bm{k})
  =
  \sum_{i,j=1}^{N_C}
  h^C_{ij}(\bm{k})
  c_{i\bm{k}}^\dagger c_{j\bm{k}},
  \end{equation}
\begin{equation}
  h^C_{ij}(\bm{k})
  =
  \varepsilon_C\delta_{ij}
  +
  \sum_{\bm{R}}
  t_{ij}(\bm{R})
  e^{i\bm{k}\cdot(\bm{r}_j+\bm{R}-\bm{r}_i)},
  \label{eq:periodic_graphene_ham}
\end{equation}
where $\bm{R}=p\bm{A}_1+q\bm{A}_2$ and $t_{ij}(\bm{R})$ includes graphene
hoppings up to sixth-nearest neighbors.

The Au and local-potential terms are
\begin{equation}
  \hat{H}_d(\bm{k})
  =
  \sum_{\alpha\in\mathcal{A}}
  \varepsilon_{\alpha}
  d_{\alpha\bm{k}}^\dagger d_{\alpha\bm{k}},
  \qquad
  \hat{H}_U(\bm{k})
  =
  U\sum_{j=1}^6
  c_{j\bm{k}}^\dagger c_{j\bm{k}},
  \label{eq:periodic_impurity_terms}
\end{equation}
where $j=1,..,6$ denote the six carbon atoms surrounding the hollow
site. In the absence of symmetry breaking, orbitals $\alpha$ and
$\bar{\alpha}$ with $m_{\bar{\alpha}}=-m_{\alpha}$ satisfy
$\varepsilon_{\bar{\alpha}}=\varepsilon_{\alpha}$.

For an ideal hollow-site impurity, the Bloch-space hybridization is written
directly in terms of the symmetry-projected Wannier coupling $t_{\alpha}$:
\begin{equation}
  \hat{H}_{\rm hyb}(\bm{k})\!\!
  =\!\!\!
  \sum_{\alpha\in\mathcal{A}}\!
  \sum_{j=1}^6
  \left[
    \frac{t_{\alpha}}{\sqrt{6}}\,
    e^{-i m_{\alpha}\theta_j}
    e^{i\bm{k}\cdot
    (\bm{r}_j+\bm{R}_j-\bm{r}_d)}
    d_{\alpha\bm{k}}^\dagger c_{j\bm{k}}
   \! +\!
    {\rm h.c.}
  \right]\!\!.
  \label{eq:periodic_hyb_second_quantized}
\end{equation}
Here $\theta_j$ is the polar angle of carbon site $j$
around the impurity center $\bm{r}_d$, and $\bm{R}_j$ is included when the
hexagon crosses a supercell boundary.

Let $\mathcal{H}_{\rm per}(\bm{k})$ denote the single-particle matrix
corresponding to Eq.~\eqref{eq:periodic_second_quantized_H}. Its retarded
Green's function is
\begin{equation}
  \mathcal{G}^R(\bm{k},\omega)
  =
  \left[
    (\omega+i\eta)\mathbb{I}
    -
    \mathcal{H}_{\rm per}(\bm{k})
  \right]^{-1}.
  \label{eq:periodic_full_green}
\end{equation}

The graphene-projected DOS per supercell is
\begin{equation}
  \rho_C(\omega)
  =
  -\frac{1}{\pi}
  \int_{\mathrm{BZ}}[d\bm{k}]\,
  \operatorname{Im}
  \sum_{i=1}^{N_C}
  \mathcal{G}^R_{c_i c_i}(\bm{k},\omega),
  \label{eq:periodic_graphene_dos_green}
\end{equation}
Here
 \begin{equation}
[d{\bf k}]
\equiv
\frac{d^2{\bf k}}{S_{\mathrm{BZ}}},
\qquad
S_{\mathrm{BZ}}
=
\frac{8\pi^2}{\sqrt{3}\,a^2},
\end{equation}
and $\rho_C(\omega)/N_C$ is the average DOS per carbon atom. The DOS
projected onto the Au orbitals with $|m_{\alpha}|=\ell$ is
\begin{equation}
  \rho_{\ell}(\omega)
  =
  -\frac{1}{\pi}
  \int_{\mathrm{BZ}}[d\bm{k}]\,
  \operatorname{Im}
  \sum_{\substack{\alpha\in\mathcal{A}\\ |m_{\alpha}|=\ell}}
  \mathcal{G}^R_{d_\alpha d_\alpha}(\bm{k},\omega),
  \qquad \ell=1,2,
  \label{eq:periodic_imp_dos_green}
\end{equation}
with
\begin{equation}
  \rho_{\rm imp}(\omega)
  =
  \rho_1(\omega)+\rho_2(\omega).
\end{equation}

To compare with ARPES and the unfolded DFT bands, we compute the
graphene-projected spectral function in the primitive graphene Brillouin
zone. A primitive momentum $\bm{q}$ is related to its folded supercell
representative $\bm{k}$ by
\begin{equation}
  \bm{q}=\bm{k}+\bm{G}_\ell ,
\end{equation}
where $\bm{G}_\ell$ is a reciprocal vector of the $5\times5$ superlattice.
For sublattice $s=A,B$, the unfolded operator is
\begin{equation}
  c_{s\bm{q}}
  =
  \frac{1}{\sqrt{N_{\rm cell}}}
  \sum_{i\in s}
  e^{-i\bm{q}\cdot\bm{r}_i}
  c_{i\bm{k}},
  \qquad
  N_{\rm cell}=25.
  \label{eq:unfolded_operator}
\end{equation}
The unfolded spectral function follows directly from the supercell Green's
function:
\begin{equation}
  A_{\rm per}(\bm{q},\omega)
  =
  -\frac{1}{\pi N_{\rm cell}}
  \operatorname{Im}
  \sum_{s=A,B}
  \sum_{i,j\in s}
  e^{-i\bm{q}\cdot(\bm{r}_i-\bm{r}_j)}
  \mathcal{G}^R_{c_i c_j}(\bm{k},\omega).
  \label{eq:periodic_spectral_function_green}
\end{equation}

\subsection{SCTMA formalism}

As noted above, to connect with experiments performed on disordered impurity configurations, we employ the SCTMA formalism to account for disorder-induced multiple scattering and to obtain the configurationally averaged electronic structure of the dilute, randomly distributed intercalant system.

For each retained Au orbital $\alpha\in\mathcal{A}$, we introduce the normalized
six-site ring vector
\begin{equation}
  \left[\bm{u}_{\alpha}\right]_j
  =
  \frac{1}{\sqrt{6}}
  e^{-i m_{\alpha}\theta_j},
  \qquad
  j=1,\ldots,6 .
  \label{eq:ring_vector_SCTMA}
\end{equation}

Integrating
out the Au orbitals gives the energy-dependent scattering potential acting
on the six neighboring carbon atoms,
\begin{equation}
  V_{\rm eff}(\omega)
  =
  U\mathbb{I}_6
  +
  \sum_{\alpha\in\mathcal{A}}
  \frac{
    |t_{\alpha}|^2
    \bm{u}_{\alpha}\bm{u}_{\alpha}^{\dagger}
  }{
    \omega-\varepsilon_{\alpha}+i\eta_{\alpha}
  } .
  \label{eq:Veff_SCTMA}
\end{equation}

In the absence
of symmetry breaking, orbitals $\alpha$ and $\bar{\alpha}$ with
$m_{\bar{\alpha}}=-m_{\alpha}$ satisfy
$\varepsilon_{\bar{\alpha}}=\varepsilon_{\alpha}$ and
$|t_{\bar{\alpha}}|=|t_{\alpha}|$. The $d_{xz}/d_{yz}$ and
$d_{x^2-y^2}/d_{xy}$ doublets then contribute through the
$m_{\alpha}=\pm1$ and $m_{\alpha}=\pm2$ ring channels, respectively.

To embed the ring-space scattering matrix into the graphene Bloch basis, we
define the $2\times6$ matrix
\begin{equation}
  \left[\mathcal{P}_{\bm{k}}\right]_{sj}
  =
  \delta_{s,s_j}
  e^{i\bm{k}\cdot\bm{r}_j},
  \qquad
  s=A,B ,
  \label{eq:P_embedding_SCTMA}
\end{equation}
where $s_j$ is the sublattice of carbon site $j$. For a random impurity
distribution of density $n_{\mathrm{imp}}=N_Au/N_C$, the SCTMA equations are
\begin{align}
  \Sigma_{\rm SCTMA}(\bm{k},\omega)
  &=
  n_{\mathrm{imp}}\,
  \mathcal{P}_{\bm{k}}
  T(\omega)
  \mathcal{P}_{\bm{k}}^{\dagger},
  \label{eq:Sigma_SCTMA}
  \\
  G_{\rm SCTMA}(\bm{k},\omega)
  &=
  \left[
    (\omega+i\eta)\mathbb{I}_2
    -
    H_0(\bm{k})
    -
    \Sigma_{\rm SCTMA}(\bm{k},\omega)
  \right]^{-1},
  \label{eq:G_SCTMA}
  \\
  g(\omega)
  &=
  \int_{\rm BZ}[d\bm{k}]\,
  \mathcal{P}_{\bm{k}}^{\dagger}
  G_{\rm SCTMA}(\bm{k},\omega)
  \mathcal{P}_{\bm{k}},
  \label{eq:g_dressed_SCTMA}
  \\
  T(\omega)
  &=
  V_{\rm eff}(\omega)
  \left[
    \mathbb{I}_6
    -
    g(\omega)V_{\rm eff}(\omega)
  \right]^{-1}.
  \label{eq:Tmatrix_SCTMA}
\end{align}
The phase factors in $\mathcal{P}_{\bm{k}}$ retain the full hollow-site form
factor, making the disorder-averaged self-energy momentum dependent. In
components,
\begin{equation}
  \left[
    \Sigma_{\rm SCTMA}(\bm{k},\omega)
  \right]_{ss'}
  =
  n_{\mathrm{imp}}
  \sum_{i,j=1}^{6}
  \delta_{s,s_i}
  \delta_{s',s_j}
  e^{i\bm{k}\cdot(\bm{r}_i-\bm{r}_j)}
  T_{ij}(\omega).
  \label{eq:Sigma_SCTMA_components}
\end{equation}

Equations~\eqref{eq:Sigma_SCTMA}--\eqref{eq:Tmatrix_SCTMA} are solved
self-consistently at each energy. Starting from $T^{(n)}(\omega)$, we
construct $\Sigma^{(n)}$, $G^{(n)}$, and $g^{(n)}$, and calculate
\begin{equation}
  \widetilde{T}^{(n+1)}(\omega)
  =
  V_{\rm eff}(\omega)
  \left[
    \mathbb{I}_6
    -
    g^{(n)}(\omega)V_{\rm eff}(\omega)
  \right]^{-1}.
  \label{eq:T_update_SCTMA}
\end{equation}
For numerical stability, the update is linearly mixed according to
\begin{equation}
  T^{(n+1)}(\omega)
  =
  (1-\lambda)T^{(n)}(\omega)
  +
  \lambda\widetilde{T}^{(n+1)}(\omega),
  \qquad
  0<\lambda\leq1 ,
  \label{eq:T_mixing_SCTMA}
\end{equation}
until
\begin{equation}
  \max_{i,j}
  \left|
    T_{ij}^{(n+1)}(\omega)-T_{ij}^{(n)}(\omega)
  \right|
  <
  \epsilon_{\rm SCTMA}.
  \label{eq:SCTMA_convergence}
\end{equation}
The converged solution at one energy may be used to initialize the next
energy. The one-shot $T$-matrix approximation is recovered by replacing the
dressed Green's function in Eq.~\eqref{eq:g_dressed_SCTMA} with the pristine
graphene Green's function. Here we consider $\lambda=0.4$ and a number of 25 iteration steps with a tolerance of $10^{-8}$.

The self-consistent feedback shifts and broadens the impurity resonances and
redistributes their spectral weight without splitting orbital partners with
opposite $m_{\alpha}$ as long as the ideal hollow-site symmetry is preserved.
The approximation sums repeated scattering from each impurity to all orders
while neglecting crossed diagrams and coherent scattering between distinct
impurities, and is therefore appropriate for dilute random intercalants~\cite{Kaasbjerg2020Tmatrix,Kot2020DefectsSCTMA}. The graphene spectral
function is
\begin{equation}
  A_{\rm SCTMA}(\bm{k},\omega)
  =
  -\frac{1}{\pi}
  \operatorname{Im}
  \operatorname{Tr}_{A,B}
  G_{\rm SCTMA}(\bm{k},\omega).
  \label{eq:A_SCTMA}
\end{equation}

\section{Pristine graphene}

\label{sec:pristine}

We first establish the reference electronic structure of pristine graphene, both from
DFT and from an effective TB description of
the graphene $\pi$ bands. This reference calculation is important since
the effect of Au intercalation will be measured against the spectral function of
undecorated graphene~\cite{NairPRB2012}.

\begin{figure}[t]
  \centering
  \includegraphics[width=1.5\columnwidth, angle=0]{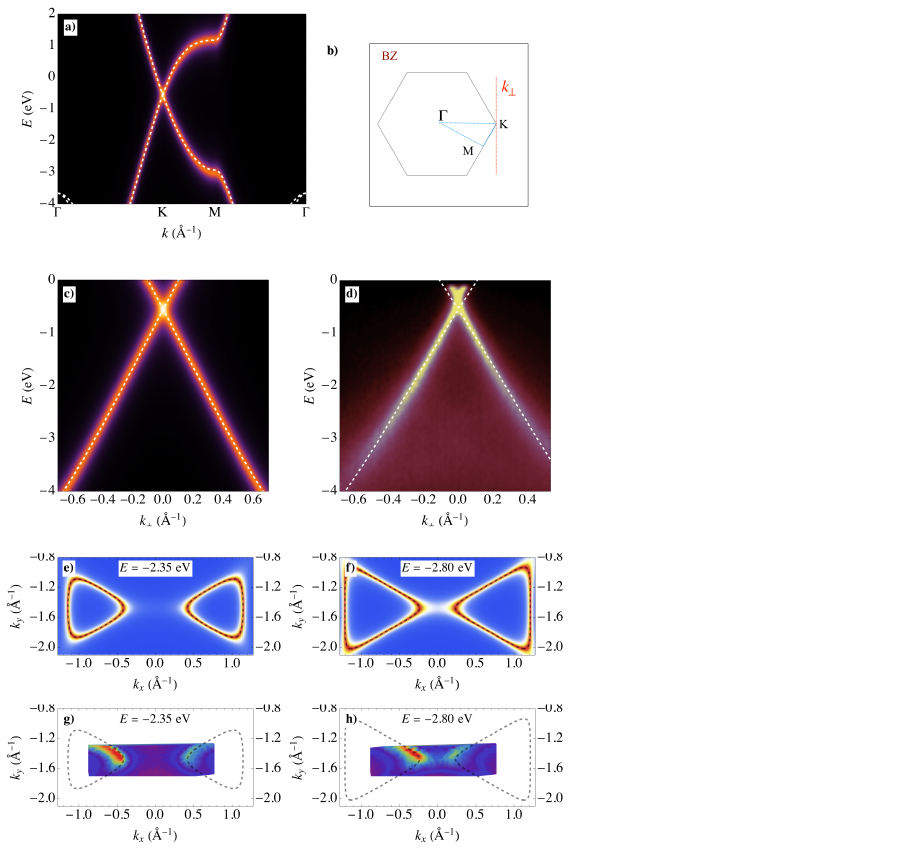}
  \caption{Band structure of pristine graphene:
  a) the TB-derived spectral function as a function of energy and momentum along the $\Gamma K M \Gamma$ path in the Brillouin zone, as indicated in b). The DFT dispersion is shown as a dashed line. c) The theoretical
  dispersion as a function of $k_\perp$, as indicated in b). The bands have been shifted by $-0.55\,{\rm eV}$ as discussed in the text. d) The corresponding ARPES measurement from Ref.~\onlinecite{NairPRB2012}, with the dashed line denoting the theoretical dispersion.  e)-h) Calculated and respectively measured constant-energy spectral
  functions at $\omega=-2.35\,{\rm eV}$ and $\omega=-2.8\,{\rm eV}$. }
  \label{fig:pristine_graphene}
\end{figure}

The DFT graphene bands are plotted as dashed lines in
Fig.~\ref{fig:pristine_graphene}. In Fig.~\ref{fig:pristine_graphene}(a) we plot the dispersion along the
$\Gamma K M \Gamma$ path in the Brillouin zone. The calculated van Hove
singularities occur at approximately $1.7\,{\rm eV}$ and $-2.38\,{\rm eV}$,
measured relative to the Fermi level used in the plot. In Fig.~\ref{fig:pristine_graphene}(c) we show
the dispersion as a function of $k_\perp$, along a line passing through $K$ and
perpendicular to the $\Gamma K$ direction. This is the geometry most directly
adapted to comparison with the ARPES measurements of Ref.~\onlinecite{NairPRB2012},
which are reproduced in Fig.~\ref{fig:pristine_graphene}(d).

By performing Wannierization and truncation of pristine graphene we find that for the energy interval relevant for experiments, the physics can be described fully by the $p_z$ orbitals, for which we need to consider in order to match the DFT dispersion, non-zero hopping amplitudes up to sixth-nearest neighbors (see Sec.~\ref{pristine} for the TB Hamiltonian and Eq.~(\ref{eq:TBparams}) for the exact values of the couplings). In Figs.~\ref{fig:pristine_graphene}(a) and (c) we plot the spectral function derived from Eq.~\eqref{eq:A0graphene} and we compare it to the DFT-derived dispersion
(denoted by the dashed lines). The agreement is quasi-perfect when the hopping terms are considered up to the 6th NN. For example the the TB  van Hove singularities occur at approximately
\[
  E_{\rm VHS}^{-} \simeq -2.47\,{\rm eV},
  \qquad
  E_{\rm VHS}^{+} \simeq 1.74\,{\rm eV},
\]
very close to the DFT predicted ones of $-2.38$eV and $1.7$eV. 

In Fig.~\ref{fig:pristine_graphene}(d)  we compare the theoretical band structure (denoted by the dashed lines) with the one measured experimentally via ARPES~\cite{NairPRB2012}. We apply a $-0.55\,{\rm eV}$ shift to the bands in order to best align the experimental data and the theory. 
This will modify the chemical potential in Eq.~\eqref{eq:H0graphene6NN} from $\mu=-377{\rm meV}$ to $\mu=173{\rm meV}$.

Note the almost perfect agreement between DFT, TB and experiments. 

To compare with the experimental observations we also plot constant-energy spectral functions at
$\omega=-2.35\,{\rm eV}$ and $\omega=-2.8\,{\rm eV}$. When taking into account the Dirac point shift of $-0.55\,{\rm eV}$, the DFT derived value for the van Hove singularity should occur in our model at $-2.93$eV. The experimental energy of $-2.8\,{\rm eV}$ is quite close to the VHS and the spectral function at $-2.8\,{\rm eV}$ consists of triangular contours that approach the saddle points near $M$, very similar to the experimental observations (see Figs.~\ref{fig:pristine_graphene}(f) and (h). On the other hand at $-2.35\,{\rm eV}$ we are quite far away from the VHS and the spectral function corresponds rather to pockets centered around the $K$ points (see Figs.~\ref{fig:pristine_graphene}(e) and (g)). The agreement
  confirms that the chosen TB parametrization provides a suitable
  effective reference model for pristine graphene in the energy range relevant to
  the Au-intercalated system.

\section{Au-cluster-intercalated graphene}

\subsection{DFT calculations}
\label{sec:dft-i}

\begin{figure}[t]
  \centering
  \includegraphics[width=\columnwidth, angle=0]{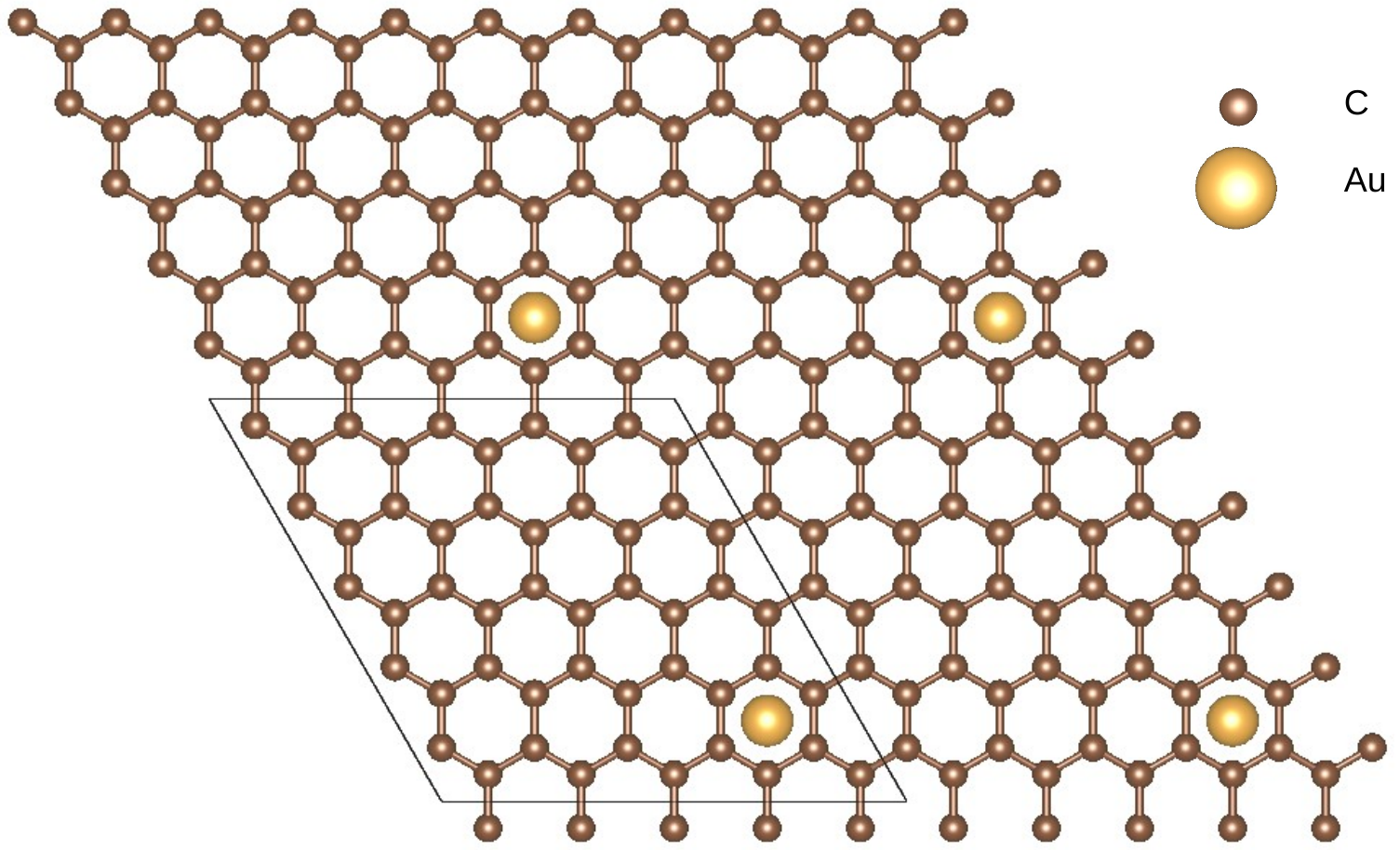}
  \caption{Representative periodic structural model used for DFT parameter extraction: uniform hollow-site Au in a $5\times5$ cell}
  \label{fig:dft_structures}
\end{figure}

Our baseline periodic approximation consists of a commensurate $5\times 5$ graphene cell with a single Au atom at the hollow position beneath the hexagon center (see Fig.~\ref{fig:dft_structures}). 

The choice of the Au atom sitting underneath the hollow site versus underneath a C atom is justified by calculating the energies of the two structures and choosing the one that minimizing the energy. Such an analysis was performed in Ref.~\onlinecite{Nair2016AuGrapheneSiC}, and we have checked that we recover the same energy minimization here. 

The structure we consider has an effective density similar to that reported experimentally for the cluster phase, which shows a characteristic approximately triangular cluster network with spacing $d\approx\SI{2.2}{nm}$ from self-correlation analysis~\cite{PremlalAPL2009}. Thus the areal density of clusters is
$n_{\mathrm{cl}}\simeq 1/\!\left(\frac{\sqrt{3}}{2}d^2\right)\approx 0.239\,\mathrm{nm}^{-2}$.
Taking three Au atoms per cluster motif, the corresponding Au areal density is
$n_{\mathrm{Au}}\approx 3n_{\mathrm{cl}}\approx 0.716\,\mathrm{nm}^{-2}$,
which corresponds to an atomic ratio Au/C $\approx 0.019$ (about 1.9\%) using the graphene areal density $n_C\simeq 38.2\,\mathrm{nm}^{-2}$.
This also provides a working density scale for SCTMA. The choice of placing Au in the hollow site, as well as the distance of $\SI{2.4}{\angstrom}$ that we use between Au and the graphene plane are consistent with the observations in Ref.~\onlinecite{Nair2016AuGrapheneSiC}. However, because the Au--graphene separation cannot be determined with absolute precision, the corresponding Au--C coupling strength can only be estimated to within its order of magnitude.

\begin{figure}[t]
  \centering
  \includegraphics[width=1\columnwidth, angle=0]{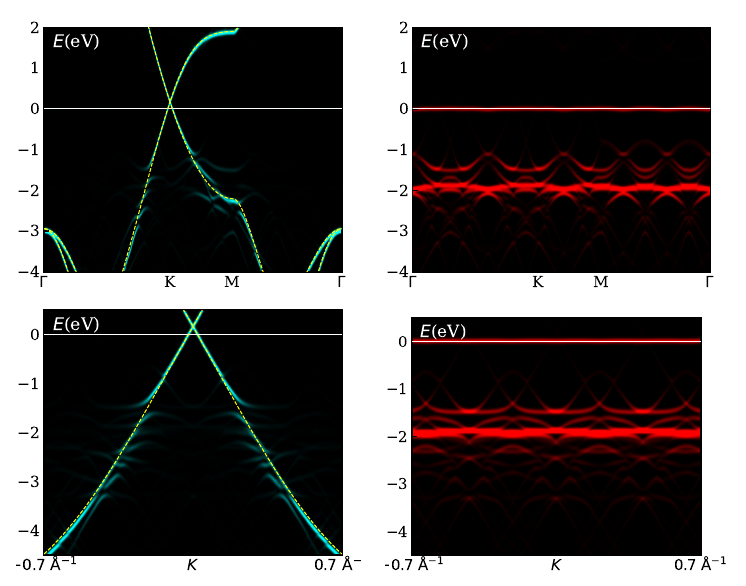}
  \caption{Band structure of graphene in the presence of a 5x5 intercalated Au from DFT calculations. On the first line the band structure plotted as a function of momentum and energy along the $\Gamma KM \Gamma$ path in the BZ, and on the second line the band structure as a function of $k_\perp$ along a direction passing through $K$ and perpendicular to $\Gamma K$ (carbon orbitals - left - in blue and Au orbitals - right - in red).}
  \label{fig:dft_unfolded}
\end{figure}

\begin{figure}[t]
  \centering
  \includegraphics[width=\columnwidth]{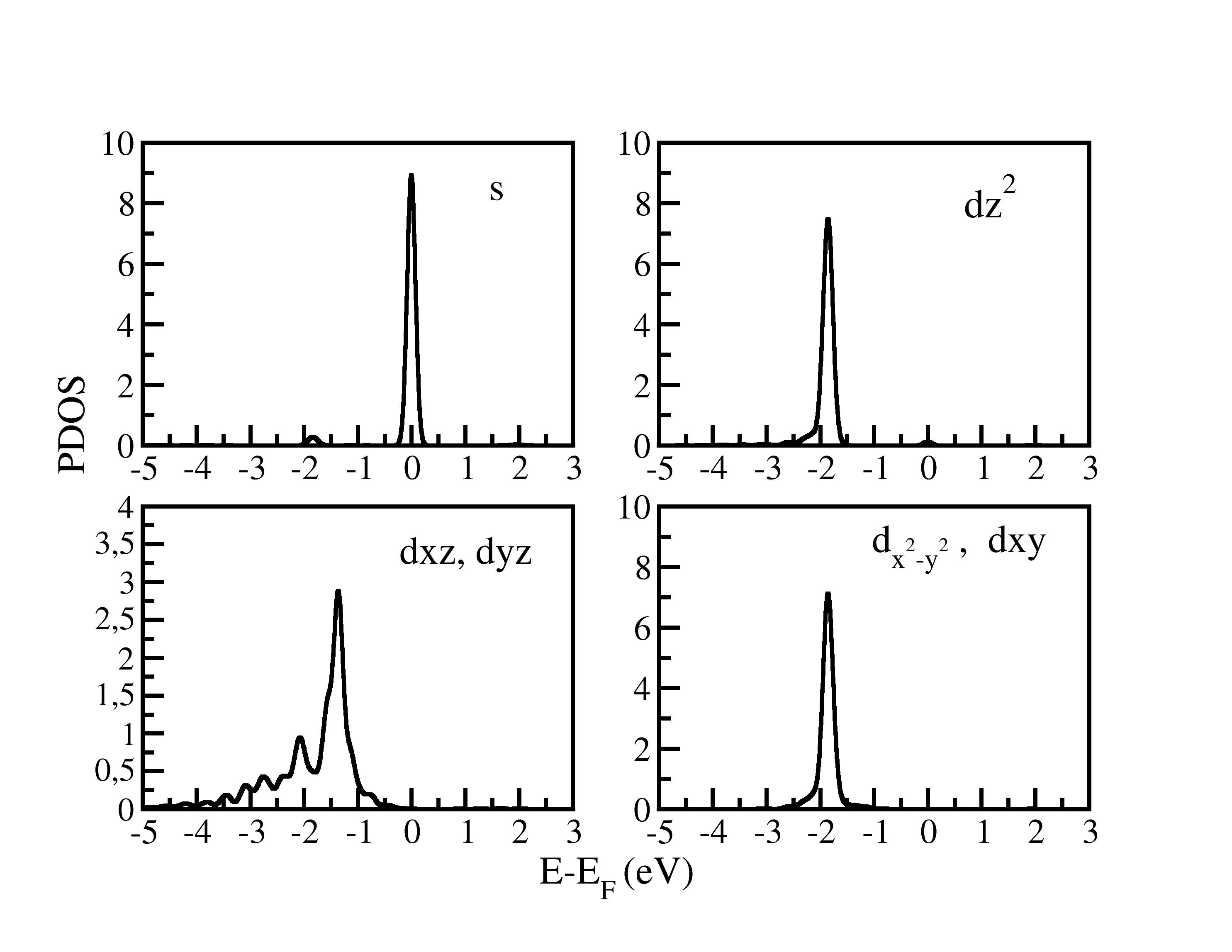}
  \caption{Orbital-resolved DOS projected onto $6s$ and $5d$ orbitals.}
  \label{fig:au_projected_dos}
\end{figure}

Fig.~\ref{fig:dft_unfolded} shows the resulting spectrum along $\Gamma K M \Gamma$ and along a $k_\perp$ cut through $K$. To identify the Au orbitals responsible for the hybridization features, Fig.~\ref{fig:au_projected_dos} shows the DOS projected onto the Au $6s$ and $5d$ orbitals. The Au $p$ states lie outside the energy window relevant for the VHS reconstruction and will not be retained in the low-energy impurity model.

The DFT spectrum shows substantial hybridization between graphene bands and Au-derived levels. In the graphene-projected spectral function (see left panels in Fig.~\ref{fig:dft_unfolded}), small gaps and deviations from the pristine dispersion appear near the intersections with Au bands. Conversely, the Au levels acquire a dispersion and broadening through hybridization with graphene (see right panels in Fig.~\ref{fig:dft_unfolded}). This effect is particularly visible for the $d_{xz}/d_{yz}$ doublet: its projected DOS is no longer a single sharp peak, but splits into several features reflecting the formation of impurity-derived bands in the periodic structure (see Fig.~\ref{fig:au_projected_dos}). Another interesting observation is that for this configuration the intercalation gives rise to a slight positive shift of the Dirac point (0.17-0.23eV QE vs OpenMX) indicating a small $p$-doping of graphene by the Au atoms; this is consistent with what has been observed experimentally~\cite{NairPRB2012}. 

A L{\"o}wdin filling analysis indicates a charge transfer of approximately 0.19$e^-$ from graphene to the Au adatom, consistent with the upward shift of the graphene Dirac point by 0.23 eV (OpenMX), indicative of $p$-type doping. This charge transfer is further supported by the integrated density of states (IDOS) of pristine graphene, which predicts a comparable carrier depletion for a Fermi-level shift of the same magnitude when scaled to the 5$\times$5 supercell. The close agreement between these independent estimates provides strong evidence that the observed Dirac-point shift arises from electron transfer from graphene to Au.

Although the resulting band structure exhibits several qualitative features, including dispersion kinks, it does not reproduce the full experimental spectrum of Refs.~\onlinecite{NairPRB2012,Nair2016AuGrapheneSiC}. As detailed below, the main sources of discrepancy are the different treatment of the intercalant distribution and the approximations inherent in the present DFT model. While the DFT calculation assumes a perfectly periodic array of Au atoms, the experimental intercalants are spatially disordered, leading to configurational averaging, momentum broadening, and scattering effects that are absent from the periodic calculation. In addition, the present model neglects several ingredients that may significantly modify the electronic structure, including the SiC substrate, the graphene buffer layer, the associated electrostatic environment and charge transfer, and spin--orbit coupling on the Au atoms.

\subsection{Wannierization, filtering and truncation}
\label{sec:wannier_ring}

The DFT analysis suggests that the modification of the graphene dispersion observed experimentally near the VHS is driven primarily by hybridization with a small number of Au-derived levels~\cite{NairPRB2012}. We therefore focus on identifying the relevant Au orbitals, their symmetry channels, and their coupling strengths to graphene.

The Wannier basis contains the graphene $p_z$ orbitals together with the Au $6s$ orbital and the full $5d$ manifold. From the resulting effective tight-binding model, we extract the Au onsite energies $\varepsilon_{\alpha}$ and the symmetry-projected hybridization amplitudes $|t_{\alpha}|$ between each Au orbital $\alpha$ and the six carbon atoms surrounding the hollow site.
The full set of extracted parameters is given in Table~\ref{tab:wannier_params}. In what follows we will not consider the $p$ orbitals since they are very far from the energy window of interest.

\begin{table}[t]
\caption{The Au--C ring parameters extracted from Wannierization. The quantity $|t_{\alpha}|$ is the symmetry-projected coupling between Au orbital $\alpha$ and the ring harmonic $c_{m_{\alpha}}$. }
\label{tab:wannier_params}
\begin{ruledtabular}
\begin{tabular}{lcccc}
Au orbital & $\varepsilon_{\alpha}$ (eV) & $|t_{\alpha}|$(eV)  & dominant channel $m_{\alpha}$ \\
\hline
$6s$ &-0.15 &  1.47 & $m_{\alpha}=0$ \\
$5d_{z^2}$ & -1.93 & 0.71 & $m_{\alpha}=0$ \\
$5d_{xz}/5d_{yz}$ &-1.86  & 0.96  & $|m_{\alpha}|=1$ \\
$5d_{x^2-y^2}/5d_{xy}$ & -1.83 &0.39    & $|m_{\alpha}|=2$ \\
$6p_z$ &6.6  &  0.25& $m_{\alpha}=0$ \\
$6p_x/6p_y$ &6.26   & 0.13 & $|m_{\alpha}|=1$ 
\end{tabular}
\end{ruledtabular}
\end{table}

The Wannier analysis shows that Au--C hoppings beyond the first carbon ring
are subdominant, with the only noticeable exception of the $d_{z^2}$-like
component. We therefore truncate the impurity potential to the six nearest carbon
atoms around the hollow site. This yields a compact ring-coupled impurity model
which retains the dominant hybridization channels while remaining directly tied
to the DFT/Wannier parameters.

To check the accuracy of the DFT/TB Wannierization and truncation,  we provide the results of a full periodic TB calculation (see Sec.~\ref{sec:periodic_tb} for details), in which graphene is coupled to a regular superstructure of Au atoms via the hopping parameters calculated using Wannierization.
Indeed, the unfolded spectral function in the TB approach, as well as the orbital projected density of states dependence on energy, closely reproduce the DFT results in the selected energy window, verifying the correctness of the DFT/TB Wannierization and truncation approach.

\begin{figure}[t]
  \centering
  \includegraphics[width=\columnwidth,angle=0]{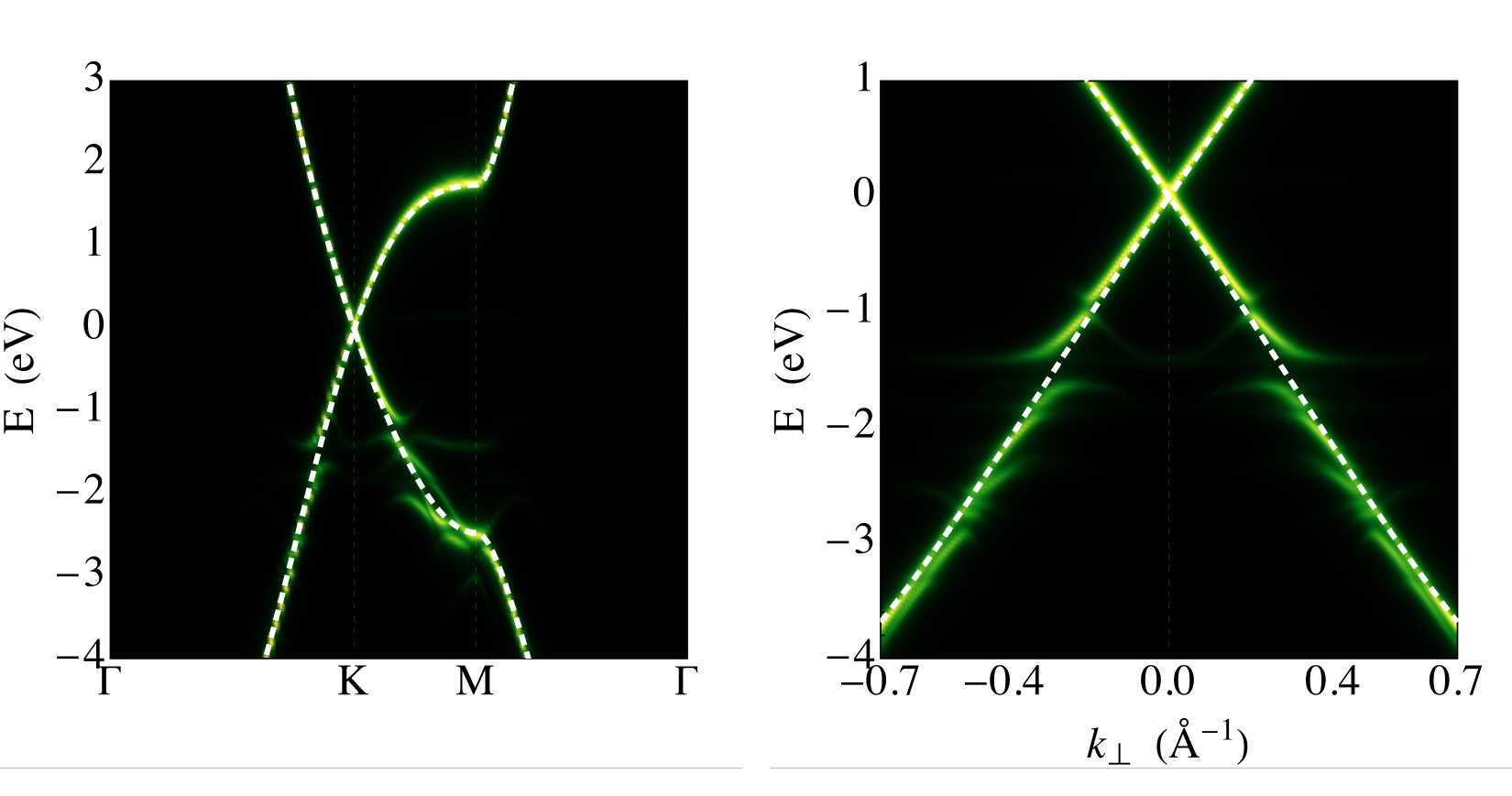}  
  \caption{The unfolded graphene-projected spectral function of the $5\times5$ Au superstructure obtained via the TB periodic formalism. The pristine graphene dispersion is denoted by the dashed line.}
  \label{fig:periodic_tb_spectral}
\end{figure}

\begin{figure}[t]
  \centering
  \includegraphics[width=\columnwidth]{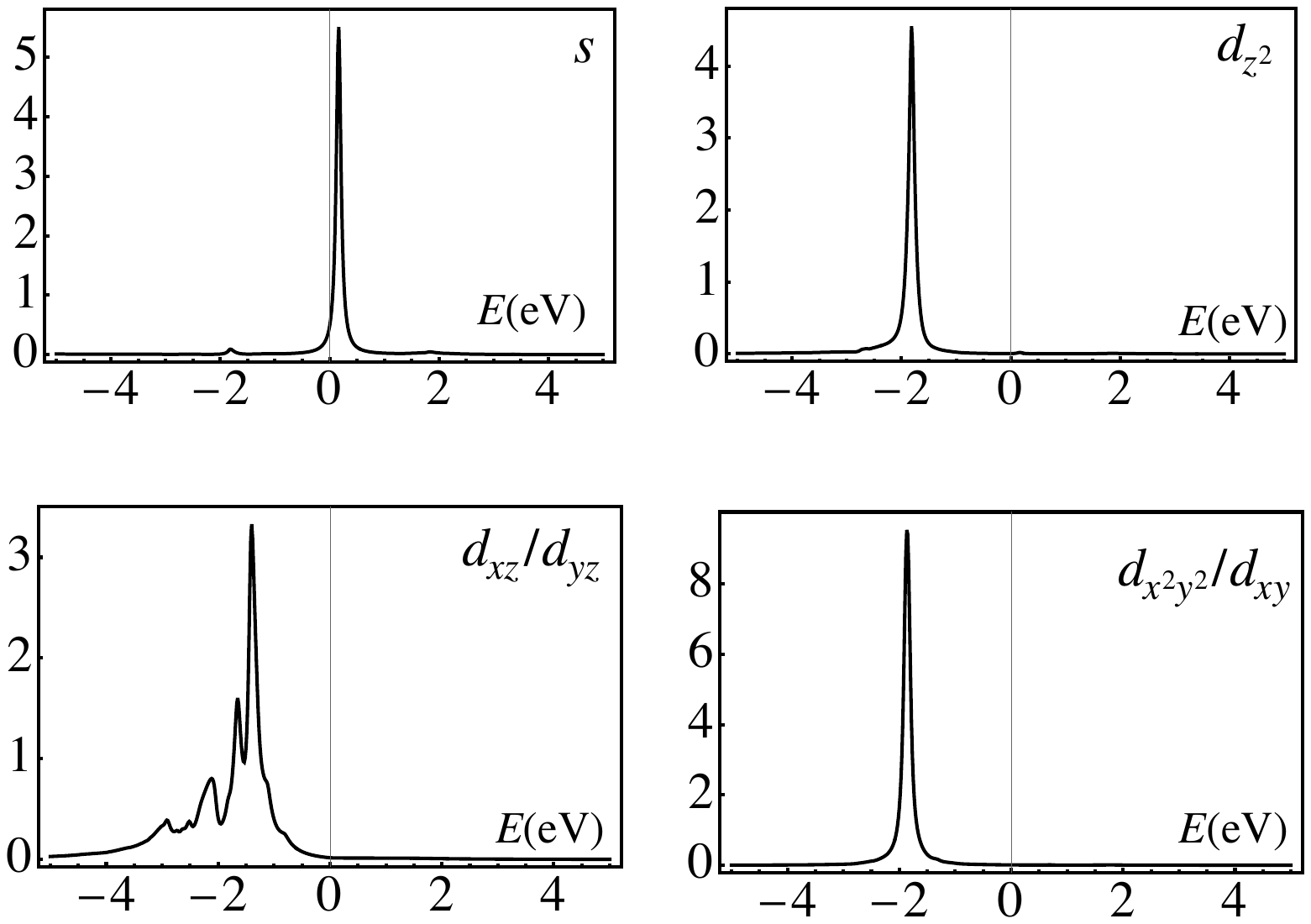}
  \caption{Density of states of the $5\times5$ Au superstructure obtained via the TB periodic formalism,
  projected onto the retained Au-derived angular channels.}
  \label{fig:periodic_tb_dos_sd}
\end{figure}

We note also that the occupied-state ARPES window, the $m_{\alpha}=0$ channel has only a minor effect on the graphene $\pi$ bands. Moreover the Au $6s$ dominant spectral weight lies close to the Dirac point. We have verified that the  $d_{z^2}$  orbital produces only a negligible modification of the band structure. We therefore reduce the parameter space further in what follows and retain only the relevant Au $5d$ orbitals $\alpha$ with $m_{\alpha}=\pm1$ and $m_{\alpha}=\pm2$.

The remaining local parameter is the static potential $U$ acting on the six carbon atoms surrounding each Au intercalant. Wannierization yields a small negative onsite shift of approximately $-0.2$~eV, although, as discussed above, this quantity is particularly sensitive to the electrostatic environment.

Direct Au--Au hopping is neglected. A separate Wannier analysis confirms that the corresponding hopping amplitudes are effectively zero at the Au separations relevant to the experiment, both between clusters, approximately $2.2$~nm, and within the clusters, where the characteristic separation corresponds to the $(2,1)$ graphene lattice vector, of magnitude $\sqrt{7},a_{\rm gr}\simeq0.65$~nm. The effective model therefore retains only the local Au--graphene hybridization channels and the static potential induced on the surrounding carbon ring.

\subsection{Effective TB model and parameters}
\label{sec:effective_tb}

As described above, the periodic DFT model that we use neglects the SiC substrate, the buffer layer, Au clustering and configurational disorder, spin--orbit coupling, and the associated electrostatic environment governing charge transfer, local carbon potentials, Au orbital energies, and the equilibrium Au--graphene separation. 
We therefore use DFT primarily as a controlled parameter-extraction tool to identify the Au orbitals and symmetry channels that hybridize with graphene and to estimate the order of magnitude of the corresponding Au--C hopping amplitudes. By contrast, the Au orbital energies, the local electrostatic potential, and the precise hybridization strengths are treated phenomenologically in the TB and SCTMA calculations and are adjusted within physically reasonable ranges to reproduce the experimental data. 

Thus DFT allows one to conclude :
\begin{enumerate}[label=\alph*)]
\item The Au atoms sit underneath a hollow graphene site
\item The Au orbitals $\alpha$ that hybridize with graphene in the experimental energy window are $5d_{xz}/5d_{yz}$ and $5d_{x^2-y^2}/5d_{xy}$ 
\item Their associated symmetry labels are $m_{\alpha}=\pm1$ and $m_{\alpha}=\pm2$, respectively
\item The hopping range and order of magnitude for the Au--C hopping amplitudes corresponds to a hybridization between Au and the nearest ring C atoms, and the hopping amplitudes between Au and C are of the order of sub-eV. 
\end{enumerate}
It cannot provide  exactly
\begin{enumerate}[resume,label=\alph*)]
\item The exact energies of the Au orbital level (due to the incomplete electrostatic environment)
\item The electrostatic potential seen by the C atoms in the vicinity of Au  (same)
\item The exact Au--C hybridization value (due to the imprecision in the Au--C distance and eventual C lattice deformations)
\end{enumerate}

For the Au--C hybridization amplitudes, we use
\begin{equation}
  |t_{\alpha}|=0.50~{\rm eV}
  \quad \text{for}  \quad d_{xz}/d_{yz},
\end{equation}
  \begin{equation}
  |t_{\alpha}|=0.25~{\rm eV}
  \quad \text{for}  \quad d_{x^2-y^2}/d_{xy}.
\end{equation}
These values are of the same order of magnitude as those obtained from the DFT/Wannier analysis, but are not expected to coincide exactly with them. The calculated couplings depend sensitively on the Au--graphene separation, which is not known with high precision experimentally. The Wannier values correspond to an Au--graphene distance of approximately $2.4$~\AA, whereas a larger separation, of order $3$~\AA, would reduce the hybridization amplitudes substantially. We have evaluated those values to to $0.4$eV for the $d_{xz}/d_{yz}$ orbitals and $0.15$eV for the $d_{x^2-y^2}/d_{xy}$ ones. Thus the values adopted here therefore lie within the physically reasonable range implied by the uncertainty in the Au--graphene distance.

For the Au orbital energies which are even more sensitive to the electrostatic environment we take
\begin{equation}
  \varepsilon_{\alpha}=-2.7~{\rm eV},
\end{equation}
for all retained orbitals.
We have verified through additional DFT calculations including spin--orbit coupling and, separately, the SiC substrate and buffer layer, that the Au $d$-orbital energies can shift towards larger binding energies by amounts of order $1$~eV. Moreover, the experimental spectral function exhibits a broad region of enhanced intensity around $-2.7$~eV. The effective Au orbital energies adopted here are therefore consistent, within the accuracy of the model, with both the extended DFT calculations and the experimental spectrum.

We use the phenomenological orbital linewidth
\begin{equation}
  \eta_{\alpha}=0.17~{\rm eV}.
\end{equation}

Similarly, the static ring potential is treated phenomenologically. The best agreement with the ARPES spectrum is obtained for $U\sim 2~{\rm eV}$, rather than the small negative onsite shift of approximately $-0.2~{\rm eV}$ extracted from the Wannier Hamiltonian. A positive $U$ corresponds, in our convention, to a repulsive short-range potential acting on electrons on the carbon ring and is consistent with the positive scattering potential inferred independently from the QPI response of the Au cluster phase~\cite{qpi}. It is also qualitatively compatible with the electrostatic effects associated with Au-induced charge transfer and the experimentally observed $p$-doping of graphene~\cite{Gierz2010Au,NairPRB2012}. We emphasize, however, that the global doping level and the local ring potential are related but distinct quantities, so the magnitude and sign of $U$ cannot be determined quantitatively from the measured charge transfer alone.

Finally, we use an impurity density
$n_{\mathrm{imp}}=2\%$, close to that inferred
experimentally. 

The DFT/Wannier estimates and the effective parameters used
in the SCTMA calculation are summarized in
Table~\ref{tab:sctma_parameters}.

\begin{table}[t]
  \centering
  \caption{DFT/Wannier estimates and effective parameters used in the
  TB and SCTMA calculations. The numerical Wannier
  hybridizations are those reported in
  Table~\ref{tab:wannier_params}.}
  \label{tab:sctma_parameters}
  \begin{tabular}{lcc}
    \hline\hline
    Parameter
    &
    DFT/Wannier
    &
    Effective value used
    \\
    \hline
    $|t_{\alpha}|$ ($d_{xz}/d_{yz}$)
    &
    0.4eV-1eV
    &
    0.5eV
    \\
    $|t_{\alpha}|$  ($d_{x^2-y^2}/d_{xy}$)
    &
    0.15eV-0.4eV    &
    0.25eV
    \\
       $\varepsilon_{\alpha}$ ($d_{xz}/d_{yz}$)
    &
  -3eV - -2eV
    &
    -2.7eV
    \\
    $\varepsilon_{\alpha}$ ($d_{x^2-y^2}/d_{xy}$)
    &
    -3eV - -2eV&
    -2.7eV
    \\
    $\eta_{\alpha}$
    &
    --
    &
    0.17eV
    \\
    $U$
    &
    -0.2eV
    &
    2eV
    \\
    $n_{\mathrm{imp}}$
    &
    $2\%$ from experiment
    &
    $2\%$
    \\
    \hline\hline
  \end{tabular}
\end{table}

\subsection{SCTMA and comparison with experiments}
\label{sec:sctma}

Having determined the effective parameters of the Au-intercalated system from DFT, we now use them in an SCTMA calculation for a random distribution of Au atoms. In the dilute limit, periodic-supercell approaches, based on either DFT or tight-binding models, and disorder-averaged methods such as SCTMA can yield comparable effective band reconstructions, provided that the impurity concentration and the local impurity--host couplings are consistently matched~\cite{HuHwangDasSarma2008,DeJuanHwangVozmediano2011,Kaasbjerg2020Tmatrix,Kot2020DefectsSCTMA}. Outside this regime, however, the two approaches may lead to different predictions. Since the experimental system is spatially disordered, it is the SCTMA results that provide the appropriate basis for comparison with the measured electronic structure.

\begin{figure}[t]
  \centering
  \includegraphics[width=1.6\columnwidth, angle=0]{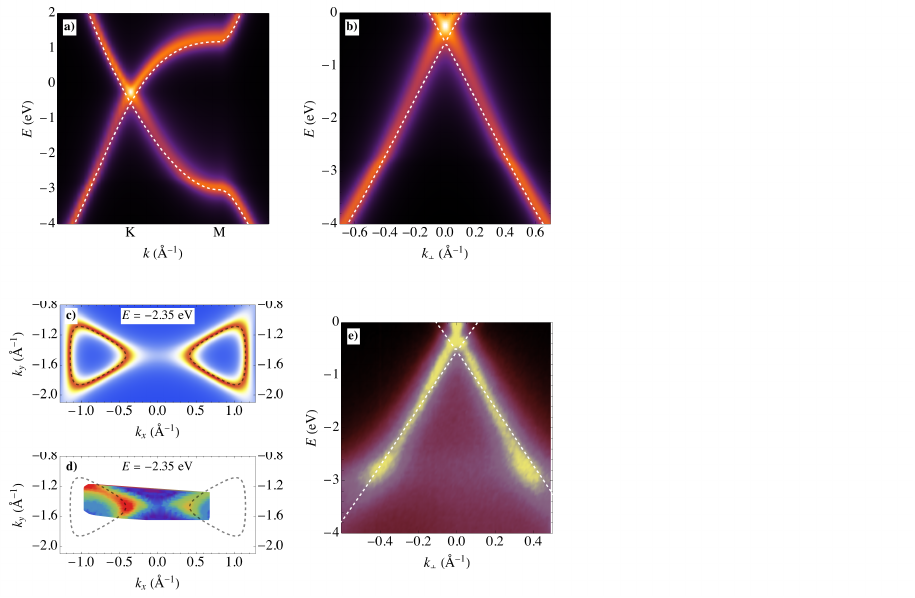}
  \caption{ a) SCTMA spectral function $A(\bm{k},\omega)$ along $\Gamma K M \Gamma$. b) and e) cuts perpendicular to $\Gamma K$ passing through $K$, theory (b) and experiment (e). 
 c) and d)  Comparison between the spectral function at $-2.35 {\rm eV}$ theory (c) and experiment (d). The unperturbed band structure is denoted by the dashed lines. The impurity levels are $-2.7 {\rm eV}$, $\mu=173$meV and $U=2$eV.}
  \label{fig:sctma_spectral}
\end{figure}

Figs.~\ref{fig:sctma_spectral}(a) and (b) show the SCTMA spectral function along the same two momentum cuts used for pristine graphene: the $\Gamma K M \Gamma$ path and a cut perpendicular to $\Gamma K$ through $K$. The comparison with ARPES in the Au-cluster (``ostrich leather'') phase~\cite{NairPRB2012} shows that the disorder-averaged impurity model captures the key phenomenology: the modification of the occupied dispersion near $M$, in particular a broadening  of the VHS. Also kink-like renormalizations near the Au-derived resonance energies $-2.7 {\rm eV}$ appear. 

In the constant-energy maps shown in Fig.~\ref{fig:sctma_spectral}(c), we use the same background linewidth, $\eta=0.17$~eV, as for pristine graphene. The contours nevertheless become broader because impurity scattering reduces the effective quasiparticle lifetime. Their radius also increases relative to the corresponding pristine-graphene contours in Fig.~\ref{fig:pristine_graphene}(e), producing a small displacement of their tips toward the $M$ points. Together, these effects generate an apparently extended VHS feature, consistent with the experimental map in Fig.~\ref{fig:sctma_spectral}(d).

The respective roles of the Au--graphene hybridization and the local ring potential $U$ can be distinguished. Hybridization with the Au-derived orbitals $\alpha$ carrying $m_{\alpha}=\pm1$ and $m_{\alpha}=\pm2$ produces a localized kink near $-2.7$~eV, close to the energy of the impurity levels, as expected for an avoided-crossing mechanism. By contrast, the effect of $U$ extends over a broader energy range: it increases the linewidth, shifts the Dirac point, modifies the energy separation between the Dirac point and the VHS, and enhances the spectral weight below the kink. It also contributes to the dome-like curvature of the bands at smaller binding energies. The presence of the same qualitative features in the experimental spectrum therefore supports the inclusion of a finite local ring potential.
\section{Conclusions and outlook}
\label{sec:concl}

We have developed a multi-level iterative DFT/TB/SCTMA/experiment method for modelling intercalated graphene systems, and applied it to the Au-cluster intercalated phase. 
The procedure consists in four steps: (i) identification of the relevant local Au orbitals via DFT; (ii) Wannierization and obtaining the order of magnitude for the values of the couplings and their symmetry channels, and incorporating them into a  TB model; (iii) a SCTMA disorder-averaging calculation of the effective spectral function; (iv) comparison with experiments in order to refine the TB parameters in the model. Our method yields the disorder-averaged momentum-resolved spectral function and thus the effective band-structure of the intercalated material.

For Au-intercalated graphene the central result is that the two main factors influencing the results are: (1) the hybridization between graphene $\pi$ bands and a small set of Au-derived $6s$ and $5d$ orbitals, in particular the $d_{xz}/d_{yz}$ orbitals with $m_{\alpha}=\pm1$ and the $d_{x^2-y^2}/d_{xy}$ orbitals with $m_{\alpha}=\pm2$, whose coupling with graphene is limited to the set of first six nearest neighbors;  (2) an onsite positive scattering potential describing the local electrostatic environment and the electron scattering from the intercalants. These two ingredients are sufficient to generate the observed modifications of the band structure near the VHS: the kink and the extension and broadening of the VHS. 

The method is broadly applicable to systems with sparse, disordered intercalants such as rare-earth, alkali, alkaline-earth, lanthanides and halide species in graphene and related van der Waals materials. It may be particularly important to model materials in which intercalation gives rise to flat bands~\cite{NairPRB2012,Rosenzweig2019Yb,Rosenzweig2020Overdoping,Zaarour2023Er}, that are believed to be relevant in the quest for high-temperature superconductivity.

\begin{acknowledgments}
We thank Catherine P\'epin, and Emile Pangburn for useful discussions and comments.
This work was supported by the ``Action amor\c{c}age du programme recherche \`a risque CEA - Audace!'' project ``Flat bands and high-temperature superconductivity in graphene''.
\end{acknowledgments}

\end{document}